\documentclass[useAMS]{mn2e}
\usepackage{epsfig,graphics}
\usepackage{amsmath}

\title[Violent intra-night variability of S4 0954+65]
{Violent intra-night optical variability of the blazar S4~0954+65 during its unprecedented 2015 February outburst}

\author[Bachev]{Rumen Bachev\\
Institute of Astronomy, Bulgarian Academy of Sciences, Sofia 1784, Bulgaria; bachevr@astro.bas.bg}

\begin{document}
\date{Accepted \dots Received \dots; in original form \dots}

\maketitle

\label{firstpage}

\begin{abstract}
In this letter we present results from intra-night monitoring in three colors ($BRI$) of the blazar S4 0954+65 during its recent (2015 February) unprecedented high state. We find violent variations on very short time scales, reaching magnitude change levels of 0.1--0.2 mag/h. On some occasions, changes of $\sim$0.1 mag are observed even within $\sim$10 min. During the night of 14.02.2015 an exponential drop of $\sim$0.7 magnitudes is detected for about 5 hours. Cross-correlation between the light curves does not reveal any detectable wavelength-dependent time delays, larger than $\sim$5 min. Color changes "bluer-when-brighter" are observed on longer time scales. Possible variability mechanisms to explain the observations are discussed and a preference to the geometrical one is given.
\end{abstract}


\section{Introduction}

Blazars are jet dominated active galactic nuclei (AGN), whose jet angles are presumably closely aligned with the line of sight. Due to the significant Doppler boosting (Doppler factors typically in the order of $\sim$20--30) relatively small variations in the jet power will appear as both larger and sharper emission variations in the observer's reference frame. Therefore, it is not surprising that many blazars occasionally manifest significant outbursts, increasing their optical emission by up to 3--5 magnitudes for a relatively short time, which is not typical for the other types of AGN. In spite of the significant progress made in the recent years and the amount of the observational work done, we are still far from fully understanding the exact nature of the processes causing blazar variability. Various mechanisms have been proposed throughout the years, from intrinsic ones (i.e. evolution of the energy density of the relativistic particles' population due to acceleration/energy losses), geometrical ones (change of the jet angle and respectively -- the Doppler boosting factor) to extrinsic ones, like microlensing from intervening bodies along the line of sight.  

One way to distinguish among these different variability mechanisms is to study the shortest time scale changes (so called intra-night variability) especially in cases where such changes are prominent. Unfortunately this happens rarely; most of the randomly selected objects for most of the time show only minor gradients or wobbles (if any) on intra-night time scales (e.g. Bachev et al. 2012 and references therein; Gaur et al. 2012; Goyal et al. 2013). On rare occasions, however, significant and well documented fluctuations of up to a magnitude for several hours have been reported (e.g. Clements et al. 2003). In this paper we study similar intra-night variability events observed recently for an interesting blazar -- S4 0954+65. These variations are among the most violent on such short time scales ever reported for a blazar.

The blazar S4~0954+65 is a $\gamma$-ray loud object (2FGL0958.6+6533) at $z\sim0.368$ with a "superluminal" jet. In the optical the object is normally rather faint with $R\sim16-17$ mag. However; starting from the end of January 2015, the optical/NIR fluxes of this blazar increased significantly (Carrasco et al. 2015), reaching (mid Feb. 2015) an $R$-band magnitude of about 13 (Spiridonova et al. 2015, Bachev 2015), a brightness level, which to the best of our knowledge has never been reported previously. Our observations reveal violent intra-night variability and cover that maximum state. It should be noted, however, that notable short-term optical fluctuations for this object are perhaps not inherent to the high optical state only. Such have been reported in other occasions at much lower flux levels (Wagner et al. 1993, Raiteri et al. 1999, Papadakis et al. 2004, Bychkova et al. 2004, Morozova et al. 2014).

In the next section we present details on the observations and data reductions. Sect. 3 presents our results, which discussed in Sect. 4. The summary of this study is given in Sect. 5.

\section{Observations and reductions}

The intra-night monitoring of S4 0954+65 was performed quasi-simultaneously in three optical bands (repeating consecutive $BRI$ frames) during four nights (the evenings of Feb. 11th to 14th) under relatively stable atmospheric conditions. The observations were done with the 60cm telescope of Belogradchik observatory, Bulgaria, equipped with an FLI PL9000 CCD camera and standard filters. The total monitoring time exceeded 20h. The integration time was 120sec with a cadence of about 8.5 min for each of the bands. Standard reduction and photometry techniques were applied to extract the magnitudes of the blazar and nearby standard stars. Stars 2, 3 and 4 from Raiteri et al. (1999) were used for the calibration of the blazar magnitudes. Star 2 (calibrated with respect to 3 and 4) was also used as a check star to additionally assess the photometric stability (Sect. 3). All magnitudes were extracted using $r=4$ arcsec diaphragm, which was reasonable for a typical seeing of 3 arcsec. The photometric errors indicated in the next section are the theoretical ones; see e.g. Bachev et al. (2005) and the references therein and Goyal et al. (2013) for possible caveats.

\section{Results}

Fig. 1 shows the $R$-band light curves for the four nights of monitoring of the blazar and a check star. Clearly, the blazar manifests violent brightness changes, reaching almost two magnitudes within the observed period of four days.

\begin{figure}
 \includegraphics[width=84mm]{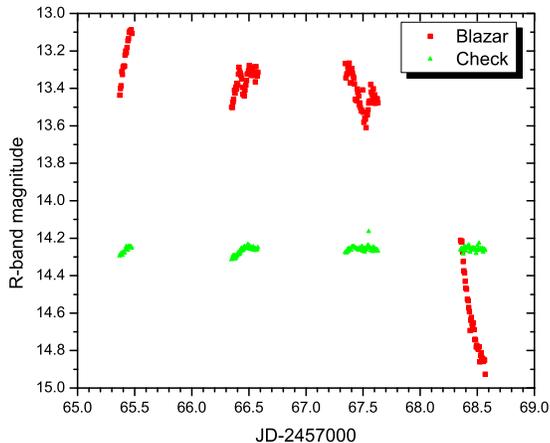}
 \caption{$R$-band light curves of the blazar and a check star during the 4 nights of monitoring}
 \label{f1}
\end{figure}

\subsection{Intra-night variations}

We show in more details the $BRI$ light curves of the blazar in Fig. 2 for the nights of February 11, 12, 13 and 14 in separate panels (from left to right and top to bottom, respectively). The blazar shows distinct intra-night variations during all four nights with typical magnitude change rates of about 0.1--0.2 mag/h. Furthermore, on some occasions (e.g. the nights of 13th; Fig. 2, bottom-left) significant frame-to-frame changes much above the photometric errors are observed in all passbands, indicating that a sampling interval of about 8.5 min might not be short enough to trace adequately the shortest-scale variations during episodes of violent activity, at least what concerns S5~0954+65. A remarkable steady decrease in brightness is seen for the night of 14th (bottom-right), where the magnitudes dropped with almost 0.7 for 5 hours.

\begin{figure*}
 \includegraphics[width=180mm]{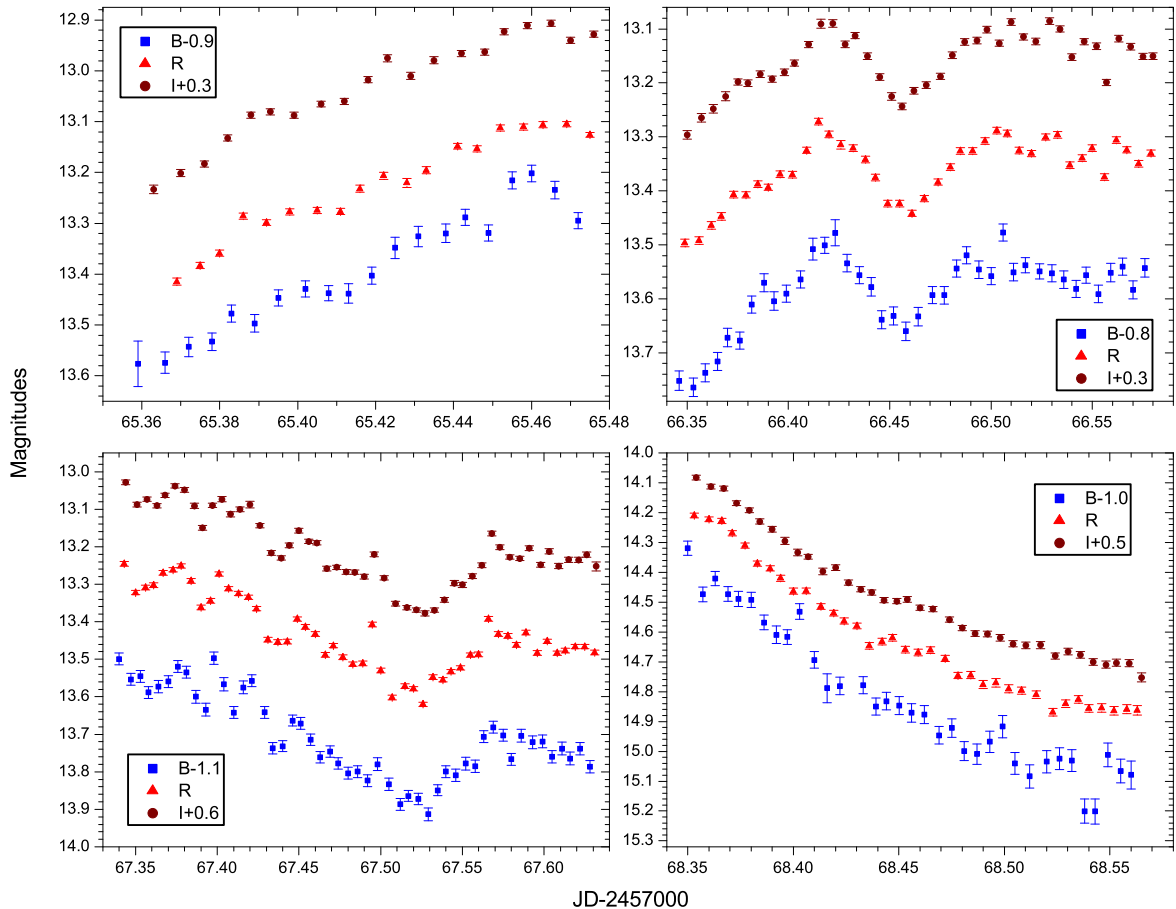}
 \caption{Individual intra-nignt $BRI$ light curves (bottom to top within a panel, respectively) of S4~0954+65 for the nights of February 11th to 14th (panels from left to right and top to bottom). Magnitude offsets are added for clarity for $B$ and $I$ bands.}
 \label{f1}
\end{figure*}

\subsection{Inter-band time delays}

In order to study any possible wavelength-dependent time delay we calculated the interpolated cross-correlation function (ICCF; Gaskell \& Sparke 1986), which gives the cross-correlation coefficient between two datasets as a function of their time displacement. This analysis has been performed for the nights of February 12th and 13th as there the light curves show prominent "structures" (i.e. multiple peaks and dips), which allows tracing the same structure from a band to the next one and thus finding the time delay between them, if such is present. Fig. 3 shows the results. Clearly, the ICCFs peak at zero lag for all bands ($B$ vs. $R$, $R$ vs. $I$ and $B$ vs. $I$), which means that there are no detectable wavelength-dependent time delays. Even though some slight asymmetry in the ICCFs towards the positive lags ("positive" means that the second place band is lagging) is present (Fig. 3, the upper panel), the results are fully consistent with a zero lag (i.e. $\tau<3-5$ min), especially taking into account the time resolution of our light curves.

\subsection{Color changes}

Multi-wavelength intra-night monitoring campaigns like the one presented in this work allow studying blazar color changes on very short timescales. Fig. 4 shows the color-magnitude ($R$ vs. $B-I$) diagram for the monitoring period. Each night is shown separately to allow distinguishing between magnitude and longer-term time dependences (if such) of the color. Overall, S5~0954+65 shows clear "bluer-when-brighter" behavior on longer time scales, which however is difficult to trace within a single night, due perhaps to the significant scatter. This scatter is partially as a result of the photometric errors (not shown for clarity) and partially -- to the fact that the $BRI$ light curve points are not exactly simultaneous (the three closest in time $BRI$ magnitudes are matched to create a point in Fig. 4). The scatter, however, should not be able to obscure entirely any chromatic behavior on intra-night time scales, if such were present. Visual inspection of the light curves in Fig. 2 also does not provide evidence for color changes within a single night.

\begin{figure}
 \includegraphics[width=84mm]{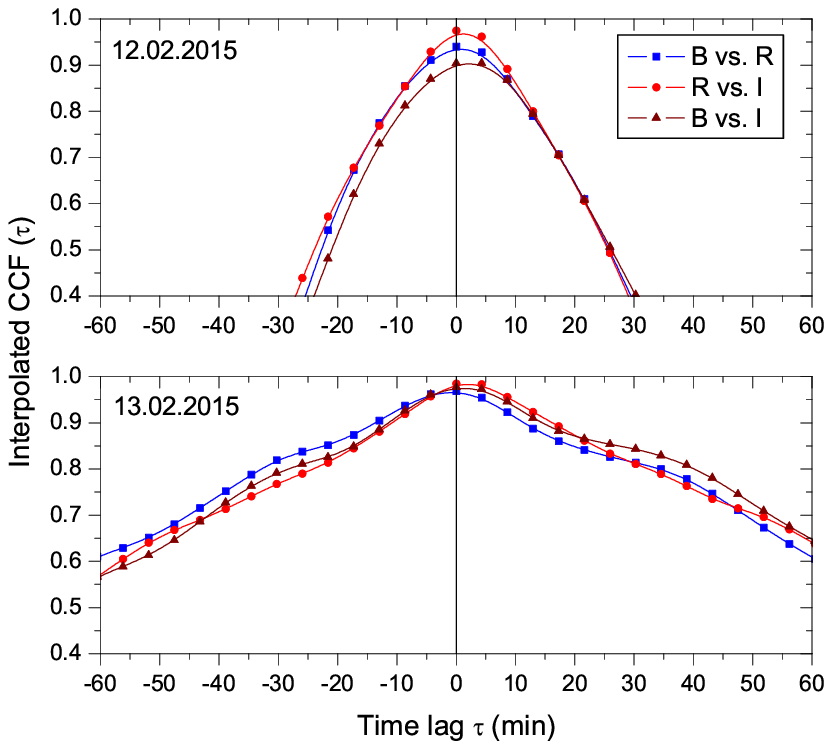}
 \caption{Interpolated CCFs for the light curves of February 12th and 13th. Peaks at zero lag indicate simultaneous variations in all bands}
 \label{f1}
\end{figure}

\begin{figure}
 \includegraphics[width=84mm]{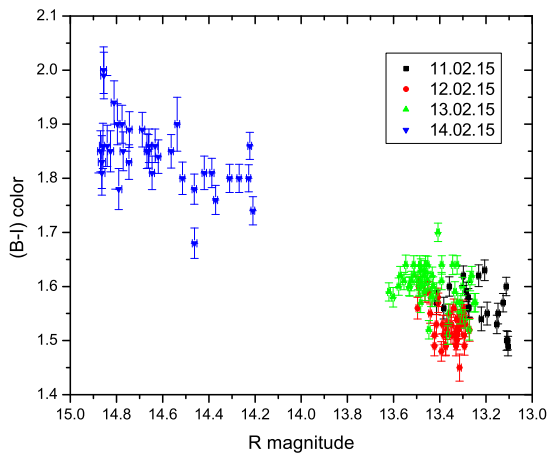}
 \caption{Brightness related color changes during the monitoring period. "Bluer-when-brighter" behavior is seen on longer time scales}
 \label{f1}
\end{figure}

\begin{figure}
 \includegraphics[width=84mm]{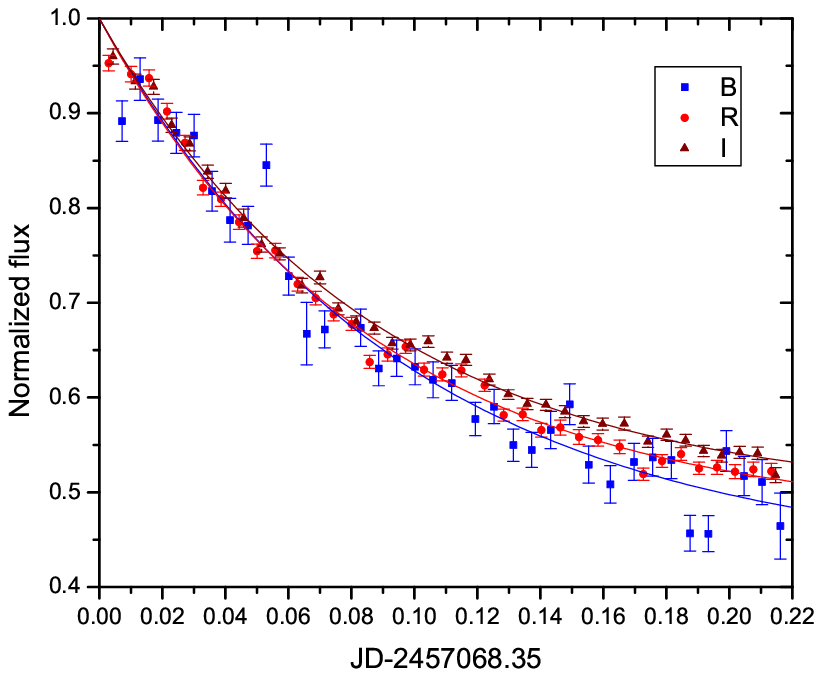}
 \caption{Exponential decay fits applied to the normalized flux data points for the huge brightness drop of 14.02.2015}
 \label{f1}
\end{figure}

Fig. 5 shows the sharp decline in brightness of the night of 14.02.2015. The smoothness of the decline allows applying a fitting procedure to the light curve points, which may be helpful to constrain the theoretical models of this and similar events. The best simple fit (among many attempted) for all bands appears to be an exponential decline in the form of $I(t)=I_{\rm 0}+I_{\rm 1}exp(-t/\tau)$ when applied to normalized fluxes. The fits are shown in Fig. 5 and the fit parameters for the different passbands are given in Table 1. Interestingly, the decay time ($\tau$) is practically the same for all pass-bands and is about 2 hours. The apparent slight chromatic behavior, considering the fits, is discussed in the next section.

\begin{table}
\begin{minipage}{84mm}
 \caption{Exponential fit parameters with uncertainties}
 \begin{tabular}{cccc}
\hline
Band & $I_{\rm 0}$ & $I_{\rm 1}$ & $\tau$ [days] \\
\hline

B & 0.427 (0.029) & 0.573 (0.025) & 0.096 (0.013)\\
R & 0.471 (0.009) & 0.529 (0.008) & 0.086 (0.004)\\
I & 0.491 (0.007) & 0.509 (0.006) & 0.087 (0.004)\\

\hline
 \end{tabular}
\end{minipage}
\end{table}

\section{Discussion}

The results presented in this work suggest that the violent intra-night variability in blazars is more likely to occur during episodes of significant outbursts (compare with e.g. Papadakis et al. 2004). Judging by our results, such outbursts, instead of being a result from a single event, consist most likely of multiple superimposed and short-living flares, resulting in large peaks and dips in the intra-night light curve of a blazar. If further confirmed for other objects, this finding can help us to better understand the mechanisms causing blazar variability. For instance, our results seem to rule out microlensing as the primary variability driver, as it most likely should be associated with a single event (a single peak, for instance). 

On the other hand the evolution of the energy density of the relativistic particles would normally be associated with some delays between the light curves as the higher energy particles evolve faster, i.e. $\dot{\gamma}\propto-\gamma^{2}$ for both -- synchrotron and inverse-Compton losses, where $\gamma$ is the Lorentz factor. We find no detectable time lags, which is not in favor of the evolution mechanism. Actually, not finding time lags is not really surprising; practically zero lags have been reported in a number of well documented studies (Wu et al. 2012 [S5 0716+714], Zhai et al. 2011 [3C 454.3], Zhai \& Wei 2012 [BL Lac], etc.). On the other hand, mostly as exceptions, lags are found on a few occasions (Papadakis et al. 2004 [S4 0954+65], Bachev et al. 2011 [3C 454.3]).

So far, the most promising mechanism to explain the intra-night variability in S5 0954+65 (but perhaps not in this object only), at least during this outburst episode appears to be the Doppler factor change of the emitting blobs, travelling down the curved or swinging jet (Gopal-Krishna \& Wiita 1992). This scenario can produce sudden brightness changes and multiple peaks (as observed) if several emitting blobs are confined closely and the direction of their movement happens to be well aligned with the line of sight at some moment. There should be no time delays between the optical bands, which is also confirmed from our results. The only drawback of this mechanism in the light of our results is the apparent color changes ("bluer-when-brighter"), evident at least at longer time scales (Fig. 4 and 5), which should be more typical if the particles' energy distribution evolves due to acceleration and/or emission losses (see Papadakis et al. 2004, for a discussion). For instance, in a case of a single injection $Q(\gamma, t)\propto\gamma^{-p}\delta(t_{\rm 0})$, the evolution as a result of the losses will always lead to redder-when-dimmer behavior, provided $p>2$ (e.g. Kembhavi \& Narlikar 1999), which is typical for blazars.

That being said, one should bear in mind that the overall blazar spectrum energy distribution (SED) in the optical region may well be a combination of the individual SED's of many contributors (emitting blobs). Thus, if the ensemble spectrum is in general redder than the spectrum of the emitting blob, whose emission is significantly enhanced at some moment due to geometrical reasons (close aligning of the velocity vector with the line of sight), than the overall bluer-when-brighter behavior can be explained within the geometrical scenario as well.

\section{Summary}

We present multi-color intra-night observations of S4~0954+65 during its unprecedented high state (2015 February). The variations we observed are among the most violent ever reported for a blazar on such time scales, reaching a change of about 0.7 mag for several hours during one of the nights. We find no detectable time delays between the optical bands. Bluer-when-brighter behavior is found, at least on longer time scales. Different variability mechanisms are discussed and arguments in support of the geometrical one are presented.

\end{document}